# High-efficiency compact optical transmitter with a total bit energy of 0.78 pJ/bit including silicon slow-light modulator and open-collector current-mode driver

KEISUKE KAWAHARA,[1,*] TAI TSUCHIZAWA,[2] NORITSUGU YAMAMOTO,[2] YURIKO MAEGAMI,[2] KOJI YAMADA,[2] SHINSUKE HARA,[3] TOSHIHIKO BABA[1]

[1]Department of Electrical and Computer Engineering, Yokohama National University, Yokohama, 240-8501, Japan
[2]National Institute of Advanced Industrial Science and Technology (AIST), Tsukuba, 305-8568, Japan
[3]National Institute of Information and Communications Technology (NICT), Koganei, 184-8795, Japan
*keisuke@ieee.org

**Increasing datacenter demands require power-efficient optical interconnects. However, a conventional standard transmitter using a silicon rib-waveguide Mach-Zehnder modulator and voltage-mode driver has low efficiency and consumes watt-class high power and occupies a several-square-millimeter footprint, which limits large-scale integration for parallel transmission. This paper presents a transmitter consisting of a compact photonic crystal waveguide (PCW) modulator and a current-mode open-collector driver. The PCW modulator is designed to have high impedance in addition to the slow-light effect. The driver connected to the modulator without termination resistors is optimized based on electronics-photonics co-simulations using a standard electronic circuit simulator with an in-house photonic model library. Co-packaging these dramatically reduces the power consumption to 50 mW and a bit energy to 0.78 pJ/bit at 64-Gbaud, and the footprint to 0.66 mm$^2$. This result represents a significant advancement toward the integration of a large number of transmission channels with no temperature control.**

## 1. Introduction

Data centers provide various internet services [1], but their power consumption is increasing rapidly and will have a significant impact on global power demand [2]. Here, optical interconnects have been deployed to reduce the power consumption of electrical wiring, where 4–16 channel optical transceivers are housed in a pluggable module to obtain high throughput [3]. The symbol rate is progressing beyond 100 Gbaud and will inevitably reach its physical limit, making parallelization an unavoidable trend. In this context, a small device footprint becomes critical, as it directly impacts the number of channels. Silicon (Si) photonics is the leading platform for integrating multichannel photonic devices [4–6]. However, Si optical transmitters, consisting of rib-waveguide optical modulators and electrical drivers, still consume significant power and footprint and make a larger number of channels difficult.

Si microring modulator (MRM) based transmitters have low power consumption and footprint [7–10]. However, because MRMs work based on resonance, they need power-hungry thermoelectric controllers (TECs) or complicated post-fabrication trimming [11] to tune the resonance. In addition, their electrical drivers are usually used in saturation mode, resulting in slow switching speeds [12,13] and limiting the symbol rate to around 50 Gbaud. On the other hand, Si Mach-Zehnder modulators (MZMs) cover a wide working spectrum without the need for TECs or trimming and achieve high-speed beyond 100 Gbaud [14–22]. However, they have a trade-off between power and footprint. To reduce the power, the drive voltage must be reduced with a long phase shifter of several millimeters. For each channel, the typical power of conventional MZMs is >100 mW, the total power of transmitters, including drivers, is >1 W, and the footprint is as large as >1 mm$^2$. To our knowledge, the most advanced study has achieved 78 mW total power at 112 Gbaud, where T-coil peaking was applied to a co-packaged Si MZM and CMOS driver, but again using long phase shifters in a footprint of 2 mm$^2$ [17].

To overcome this issue, we have studied Si photonic crystal waveguide (PCW) MZMs [23]. The slow-light effect of the PCW enhances the modulation efficiency in proportion to the group index $n_g$, enabling a phase shifter length of less than 200 μm. By engineering EO phase matching, this modulator demonstrated 64 Gbaud operation [24]. The driving voltage $V_{pp}$ has been reduced up to 0.87 V, and the power consumption and corresponding bit energy, excluding drivers, have also been reduced to 3.8 mW and 59 fJ/bit, respectively, due to electrical impedance engineering [25]. Therefore, the next target we focused on in this study is the reduction of total power, including drivers.

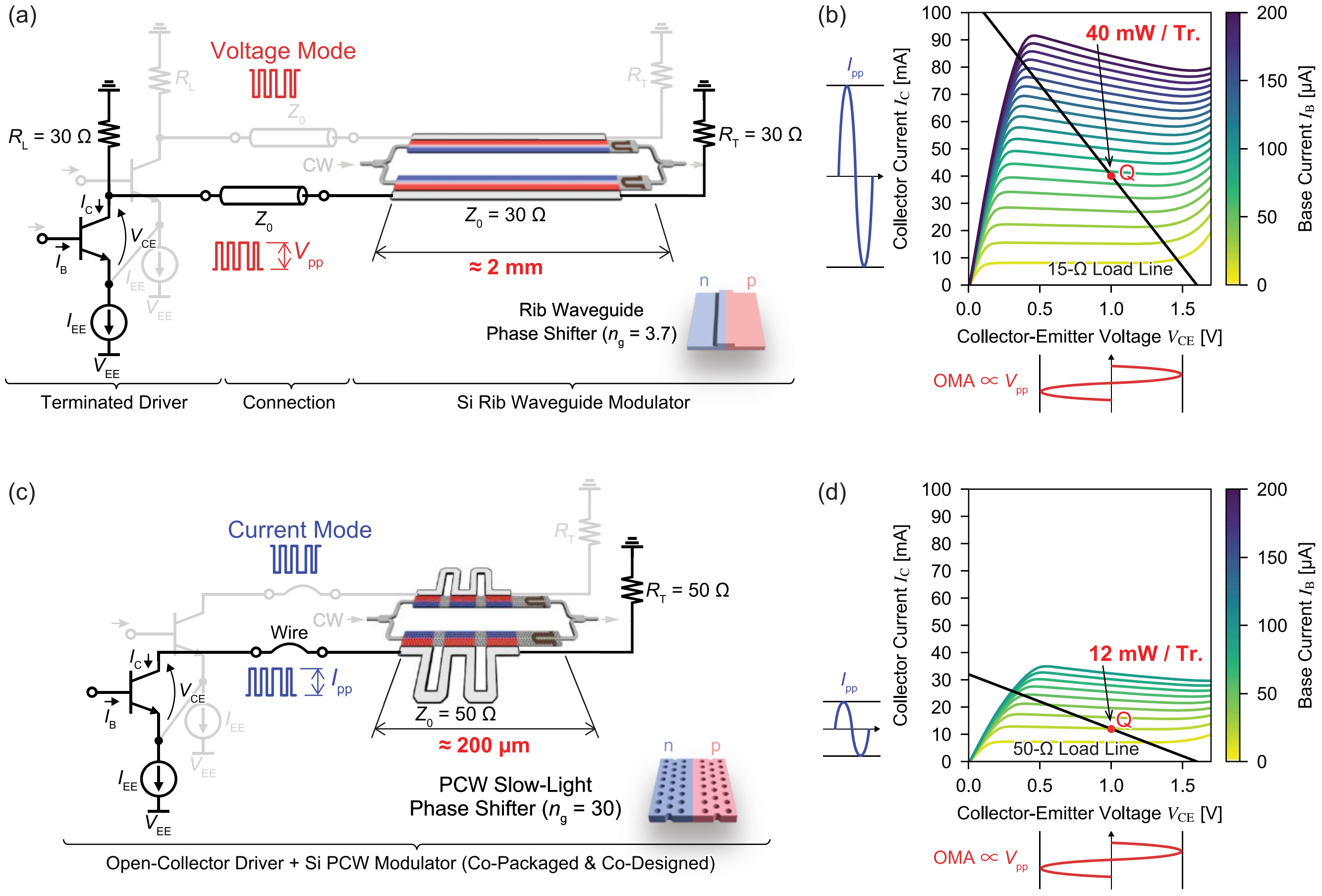

Fig. 1. Frontend configuration of an optical transmitter using a MZM. (a) Traditional driving for a rib-waveguide phase shifter with a voltage-mode driver using termination resistors. (b) Static characteristics and load line of the transistor in (a). (c) Proposed open-collector current-mode driving for a high-impedance Si PCW MZM. (d) Static characteristics and load line of the transistor in (c).

As understood from the power values mentioned above, the driver power is dominant in transmitters, although it is reduced in proportion to the modulator power. So far, most studies have employed voltage-mode drivers with termination resistors for electrical impedance matching. However, the termination resistors at drivers and modulator phase shifters doubly consume power. To solve this problem, we employed open-collector current-mode drivers with no termination resistors in this study. While reducing power, a lack of termination resistors and impedance matching requires the MZM and drivers to be integrated in close proximity. Furthermore, it severely complicates the optimization of drivers; open-collector drivers reported so far only achieved low symbol rates and efficiencies [26,27]. Therefore, we developed an in-house photonic device library that can be used in a de facto standard electronic circuit simulator and enabled the co-simulation and co-design of the MZM and drivers [28]. We finally fabricated the transmitter and successfully demonstrated record-level low power and small footprint.

In this paper, we next show the device concept and driver architecture in more detail. We explain our co-simulation method and optimized structures with expected performance, and then present the experimental results.

## 2. Design Concept

Figure 1(a) shows a typical frontend configuration of a transmitter using rib-waveguide MZMs and terminated drivers. The phase shifter is several millimeters long, resulting in a large junction capacitor as well as a large footprint. Typically, the capacitor reduces the characteristic impedance of the electrodes, $Z_0$, to ~30 Ω [29], and both termination resistors $R_T$ and $R_L$ must be close to 30 Ω for impedance matching. As a result, the combined impedance seen from the transistor becomes as small as 15 Ω. In high-speed MZMs, the p-n junction is reverse-biased, and the phase shift is induced by the voltage. Thus, a large driving current $I_{pp}$ flows in the modulation. As shown in Fig. 1(b), to obtain a drive voltage of $V_{pp}$ = 1 V for a 15-Ω load, each transistor dissipates 40 mW of DC power at the operating point. Furthermore, the parasitic capacitance increases due to the multi-parallel transistors that supply large currents, which limits the bandwidth. Thus, the conventional configuration has challenges in terms of power consumption, operating speed, and footprint.

To address these challenges, we propose a transmitter configuration shown in Fig. 1(c), including a Si PCW MZM with high-impedance electrodes and an open-collector current-mode driver. The PCW phase shifter of shorter than 200 μm results in a small footprint and a small capacitor, and the impedance can be increased to ~50 Ω in combination with moderately large inductors of the meander-line electrodes for EO phase matching. This high-impedance MZM is operated with the driver without the load resistance for the transistor, $R_L$. Then, this makes $R_T = 50\ \Omega$, and only 20 mA is required to obtain a driving voltage of $V_{pp} = 1$ V, as shown in Fig. 1(d). The phase shift, $\Delta\varphi$, is induced in proportion to the voltage, thus reducing power consumption to (12 mW/40 mW) = 30% while maintaining the modulation amplitude. In addition, the parasitic capacitance of the transistor is reduced, making it easier to increase the bandwidth of the driver. Consequently, the proposed configuration provides low power consumption, high speed, and a small footprint. Such a design concept is valid independently of the specific materials, fabrication process, and implementation. Therefore, we can apply it not only to Si photonics and BiCMOS electronics but also to thin-film lithium niobate [30] and other material high-speed modulators. Drivers can be fabricated using CMOS [31,32] or III-V semiconductors [33]. Wire bonding and flip chip bonding are available for implementation, or the monolithic integration of photonics and electronics further reduces parasitics, although the process becomes more complicated [19].

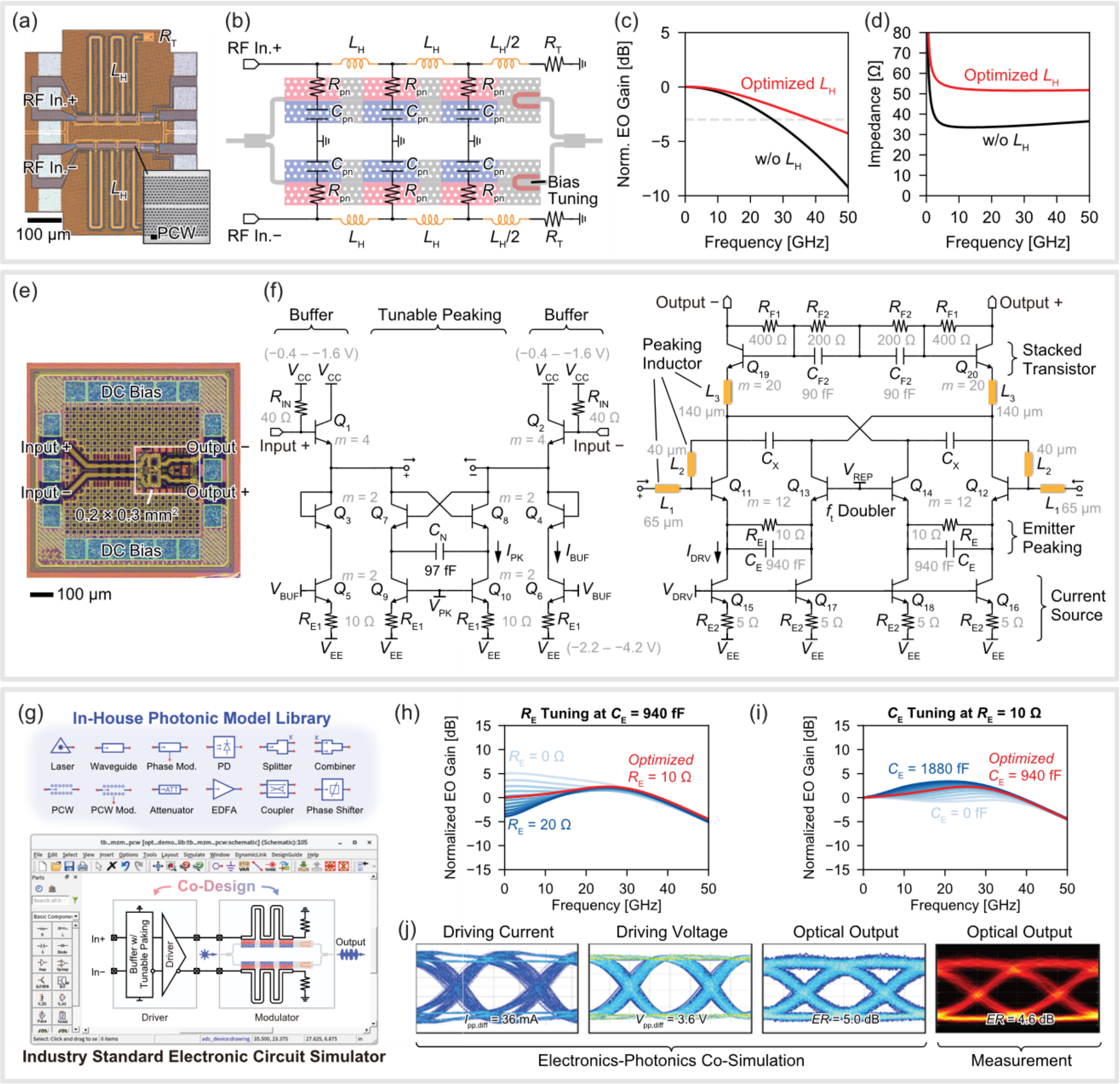

Fig. 2. Electronics-photonics co-design of the transmitter. (a) Micrograph of the modulator. (b) Circuit model of the modulator electrode. (c) Simulated EO gain of the modulator. (d) Simulated characteristic impedance of the electrode. (e) Micrograph of the modulator. (f) Schematic circuit diagram of the driver. (g) Co-simulated EO gain of the whole transmitter, including the driver and modulator for various emitter resistors. (h) Co-simulated EO gain of the whole transmitter, including the driver and modulator, for various emitter capacitors. (j) Prediction of unprocessed eye diagrams at 50 Gbps through co-simulation.

The challenge in employing an open-collector current-mode driver is that the lack of impedance matching may cause RF reflections, which degrade signal quality. It is difficult to properly predict and control RF reflections with the conventional approach that designs and packages the MZM and drivers separately.

Therefore, we have introduced electronics-photonics co-simulation in a single platform. To import photonic devices into standard electronic circuit simulators, we developed an in-house photonic model library written in the hardware description language Verilog-A [28]. The library includes fundamental photonic components such as lasers, waveguides, photodetectors, couplers, optical amplifiers, and phase shifters. Each model is accurately modeled up to 64 Gbaud based on device simulation and measurements. In this study, we used Keysight ADS with this library.

## 3. Electronics-Photonics Co-Design

Figures 2(a) and 2(b) show a micrograph of the Si PCW MZM and a circuit model of the electrode. The design of the MZM is the same as that reported previously [25]. The phase shifter length was only 150 μm, and the footprint was reduced to 0.3 mm² including the electrodes. The p-n doped phase shifters were divided into 50-μm-long segments, and inductors $L_H$ were inserted between them, where the characteristic impedance $Z_0 = \sqrt{(L_H/C_{pn})}$. Figure 2(c) shows the co-simulated EO frequency response. Ensuring EO phase matching, the 3-dB bandwidth of the response was extended up to 39 GHz. Figure 2(d) shows the simulated impedance, showing that the $L_H$ increases $Z_0$ from 30 to 50 Ω, which reduces the current flow.

Figure 2(e) shows a micrograph of the driver. The active area of the driver is only 0.2 × 0.3 mm², thanks to the compact device size enabled by the current-mode design. The driver consists of two stages (buffer and driver), as shown in Fig. 2(f). Multiple peaking techniques were implemented in the driver to compensate for the modulator's loss and extend the bandwidth. $Q_{11}$–$Q_{14}$ form a $f_t$-doubler, reducing the input capacitance of the driver stage while maintaining sufficient output current. Capacitor $C_X$ neutralizes the collector-emitter parasitic capacitance of $Q_{11}$ and $Q_{12}$ to ensure stability. The driver stage utilizes two stacked transistors, $Q_{19}$ and $Q_{20}$, to achieve a high breakdown voltage. The feedback circuits, consisting of $R_{F1}$, $R_{F2}$, and $C_{F2}$, ensure voltage balance between the stacked transistors. Collector-emitter diodes are used for $C_{F2}$ and $C_X$ to ensure device matching. $L_1$ to $L_3$ are peaking inductors that enhance high-frequency gain by resonating with parasitic capacitance. The buffer stage incorporates a tunable peaking circuit based on a negative impedance converter ($Q_7$ and $Q_8$) using positive feedback, compensating for frequency variations caused by manufacturing variations, temperature fluctuations, and power supply voltage changes. In this design, the termination of the modulator is connected to the ground, so the driver operates with a negative bias. When applying a similar design to a CMOS process, the substrate of the CMOS chip must be at the lowest potential within the chip. If this voltage is defined as ground, a positive bias needs to be applied to the modulator's termination. In CMOS design, the low breakdown voltage becomes a challenge, which can be addressed using a stacked-FETs circuit with triple-well MOSFETs [34]. Therefore, a methodology similar to this study can also be applied to CMOS drivers.

To predict the overall response of the transmitter, all driver device parameters were optimized through the co-simulation shown in Fig. 2(g), including the drivers, MZM, and bonding wires. Figures 2(h) and 2(i) show the EO frequency response of the entire transmitter when varying $R_E$ and $C_E$ as representative examples. Sufficient bandwidth cannot be obtained for small $R_E$ or large $C_E$, whereas excessive peaking is observed for large $R_E$ or small $C_E$. Balanced performance is expected for $R_E$ = 10 Ω and $C_E$ = 940 fF, and achieved $f_{3dB}$ = 46 GHz.

Other circuit parameters, such as transistor dimensions, bias currents, and all other resistors and capacitors, were fully optimized. Figure 2(j) shows the 50-Gbaud eye diagrams obtained through simulation. Despite the complexities of the current-mode design, the simulation accurately predicted the measurements.

## 4. Results and Discussion

### A. Device Fabrication

Figures 3(a) and 3(b) show photographs of the co-packaged transmitter. The MZM was fabricated using the 300-mm Si-on-insulator process of Advanced Industrial Science and Technology, and the driver was fabricated using the SG13G2 130-nm BiCMOS process of Leibniz Institute for High Performance Microelectronics (IHP). These two chips were bonded on a printed circuit board (PCB) and connected with aluminum wires. Although several devices were integrated into the Si photonics chip, only one of them was used. The active footprint, including the MZM, driver, and wires, was only 0.66 mm². Figure 3(c) shows the cross-section of the transmitter. The PCB under the modulator was cavitated, and the wires were shortened.

To investigate the impact of thermal coupling in co-packaged devices, Fig. 3(d) presents temperature simulation for power dissipation ranging from $P_{DC}$ = 100 mW to 300 mW using Autodesk Fusion. The results indicated that, at $P_{DC}$ = 300 mW, the PCW phase shifter was heated up to 84°C. Figure 3(e) shows the optical transmission spectrum of the PCW MZM for $P_{DC}$ = 100–300 mW. The on-chip insertion loss was approximately 11 dB, including the p-n doped PCW phase shifter loss of ~8 dB. The PCW in this design has a 3-dB operating bandwidth of 5.9 nm, whereas the wavelength shift between 0 mW (25°C) and 300 mW (84°C) was only 2.6 nm, meaning that the device can operate without laser wavelength tuning or active temperature control. Figure 3(f) presents the measured spectrum of the group index $n_g$ of the PCW. The PCW in this design exhibits wavelength dispersion with $n_g$ = 20–70. Even assuming a ±1 nm spectral shift around the wavelength of 1537 nm, the variation in $n_g$ is limited to ±5. Although slight variations in $n_g$ affect the frequency response of the MZM, this is evaluated later through frequency response measurements. Furthermore, well-optimized PCW structures can show a wider operating bandwidth and suppress wavelength dispersion. We have calculated that $n_g$ of approximately 20 and 30 are available with a wavelength bandwidth of 35 and 20 nm, respectively [35]. Applying such structures will lead to the development of even more thermally robust transmitters [36]. To maximize the modulation efficiency, the initial phase difference between the two arms was set at $\pi/2$ (quadrature bias). As shown in Fig. 3(g), the bias was controlled by heating the PCW using an on-chip TiN heater. Figure 3(h) shows the normalized output power of the MZM for the heating power. The required power for $\pi/2$ bias was at most 3.3 mW, which is sufficiently small compared with the total power consumption.

### B. Frequency Response

Figures 4(a) and 4(b) show a block diagram of the frequency response measurement setup and the probing details. Continuous-wave laser light was input into the MZM, and a sinusoidal signal from a signal generator was applied to the driver. The modulated optical signal was amplified, filtered, and observed using an optical sampling oscilloscope. The losses from the input RF cables and probes were de-embedded after measurement. For more details on the experiment, refer to Appendix C. Figure 4(c) presents the measured EO frequency response of the transmitter at $n_g = 35 \pm 5$, along with the co-simulation results. In this measurement, the driver's supply voltage was set to $V_{EE} = -4.2$V, and the power consumption was $P_{DC} = 260$ mW. The measurements showed good agreement with the co-simulation under all test conditions. As $n_g$ increased, the EO phase mismatch of the MZM became larger, leading to bandwidth degradation. However, tunable peaking maintained $f_{3dB} = 45$ GHz even at $n_g = 40$. Under the condition of $n_g = 30$ with the peaking, the widest bandwidth of $f_{3dB} = 55$ GHz was recorded, suggesting the potential for operation beyond 100 Gbaud.

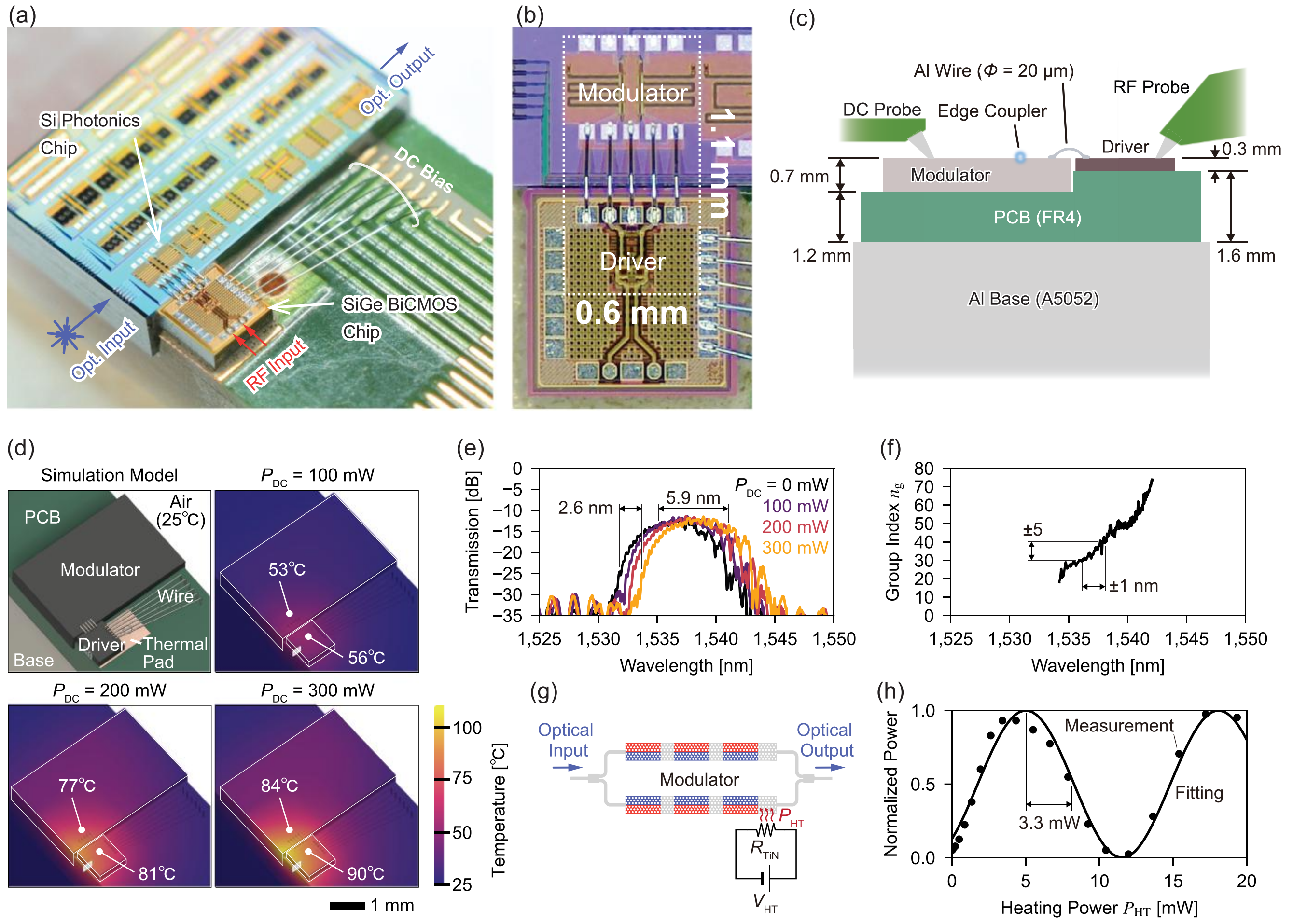

Fig. 3. Integrated optical transmitter including a Si PCW MZM and current-mode driver. (a) Bird's eye photograph (b) Top view micrograph. (c) Cross section schematic. (d) Temperature simulation for various power consumptions to investigate the thermal crosstalk. (e) Measured optical transmission spectrum of the transmitter. (f) Measured $n_g$ spectrum of the PCW. (g) Schematic of MZM bias adjustment through on-chip heating. (h) Measured normalized optical power versus heating power.

### C. Transmission Performance

The performance of the transmitter was evaluated through signal transmission experiments with a setup shown in Fig. 5(a), which is similar to the frequency response measurement; a pseudo-random non-return-to-zero (NRZ) signal was generated using a pulse pattern generator and multiplexer. Modulated light was amplified and filtered before observing with an optical sampling oscilloscope. After the experiments, digital signal processing (DSP), including symbol synchronization and 5-tap linear feed-forward equalization, was applied at the receiver, while no DSP was applied at the transmitter. The bit error rate (BER) was estimated by fitting the vertical histogram at the decision point of the eye diagram with two Gaussian functions. See Appendix D for details.

Figures 5(b) and 5(c) show the unprocessed and equalized eye diagrams at data rates of 50 and 64 Gbps for different bias power levels ranging from $P_{DC} = 50$–$260$ mW. The bit energy is shown for each case, with the extinction ratio (*ER*) and the corresponding forward error correction (FEC) compatibility. The corresponding BERs are summarized in Fig. 5(d). At 50 Gbps, the transmitter demonstrates a clear unprocessed eye opening with HD-FEC compatibility at all tested power levels, with *ER* increasing from 1.5 dB at 50 mW to 4.6 dB at 260 mW. Even at the maximum power level, the power consumption remains only one-fourth of that of conventional watt-class MZM drivers. At 64 Gbps, the eye opening was slightly degraded but still maintained SD-FEC compatibility down to $P_{DC} = 70$ mW. The best bit energy in DSP-free operation reached 1.0 pJ/bit. Applying a receiver equalizer effectively improved the eye opening, clearing the FEC threshold at 64 Gbps for all power levels with the lowest bit efficiency of 0.78 pJ/bit for a power consumption of 50 mW. In this measurement, the data rate was limited to a maximum of 64 Gbps by the multiplexer. However, given the remaining margin in the 64 Gbps eye opening and the small-signal response approaching 50 GHz, the transmitter exhibits the potential for a higher data rate. Furthermore, by employing advanced packaging techniques such as flip-chip bonding, future extensions beyond 100 Gbps per lane can be achieved while maintaining power consumption and potentially improving bit energy to less than 0.5 pJ/bit. Figure 5(e) illustrates the power consumption breakdown of the transmitter under two different bias conditions, with $P_{DC}$ = 260 mW and 50 mW. In the high-power condition, most of the power is consumed by the termination resistors and the open-collector driver stage. In the low-power condition, the main contributors to power consumption are the open-collector driver transistors, but a small amount of additional power is required for the peaking circuit to compensate for bandwidth limitation. Beyond these components, additional power consumption arises from the laser (80 mW under assumed 20% wall-plug efficiency) and from the bias control of the MZM (3.3 mW at maximum). Including these components, the total power consumption on the transmitter side might increase to 133 mW, and the bit energy is 2.1 pJ/bit.

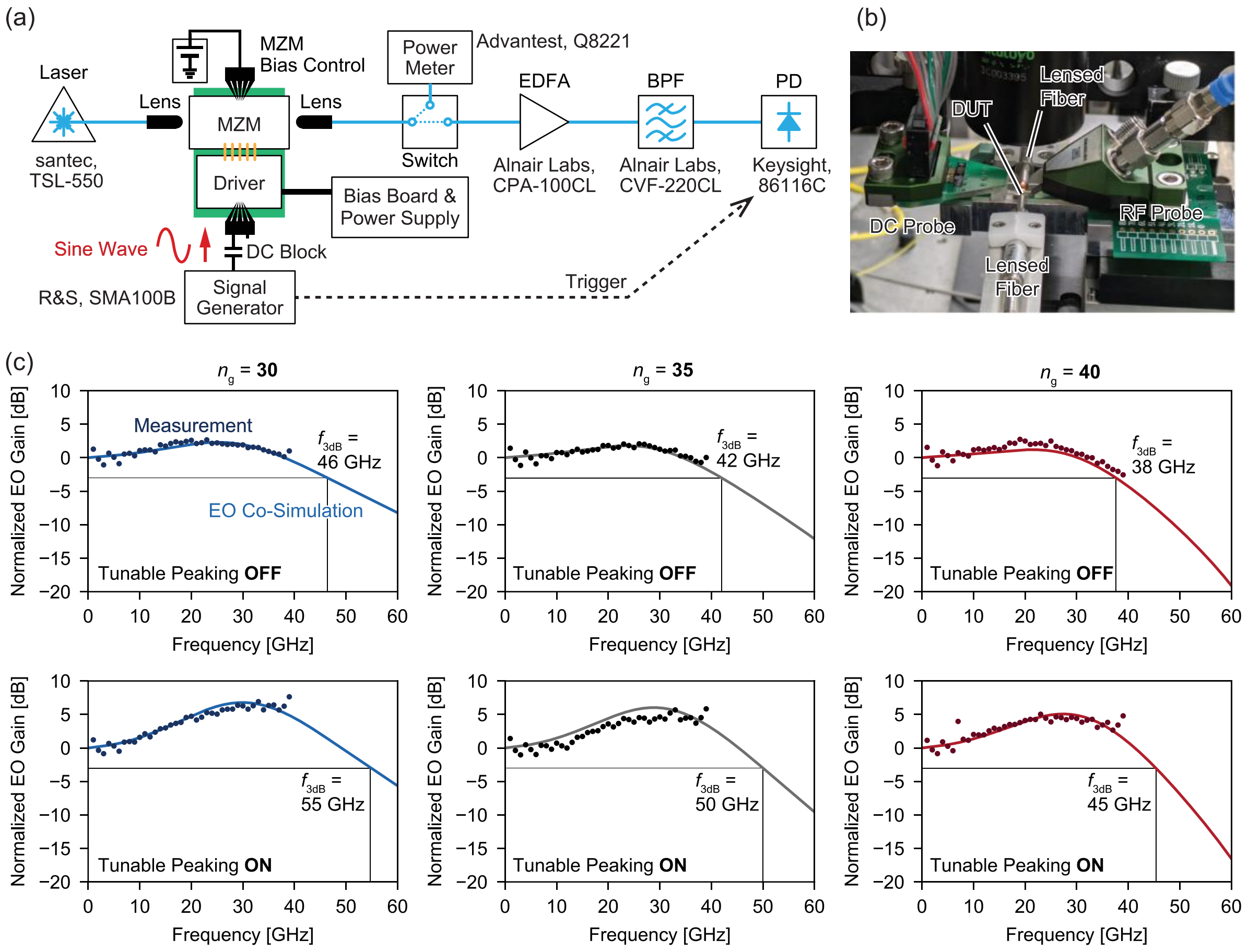

Fig. 4. Measurement of transmitter's EO frequency responses. (a) Block diagram of the experimental setup. (b) Probing details. (c) Measured and co-simulated responses at $P_{DC}$ = 260 mW for $n_g$ = 35 ± 5.

Table 1 lists the state-of-the-art EO-integrated Si optical transmitters. The traditional terminated driver underutilized the transistor current, reducing power dissipation significantly and sacrificing modulation amplitude, and the long phase shifter increased the footprint to 2 mm$^2$ [17]. A 16-channel transmitter was demonstrated using a Si modulator and CMOS driver [14], but the long rib-waveguide MZM resulted in high power consumption and footprint. Monolithic integration processes are inferior to others in terms of data rate, power consumption, and footprint due to the limited performance of electronic devices [14,19]. Electro-absorption modulators (EAM) have demonstrated an excellent efficiency and footprint, but the drawbacks are CMOS compatibility and nonlinearities due to carrier saturation [37]. Our transmitter achieves a footprint of 0.66 mm$^2$, which is significantly smaller than other MZM-based transmitters and approaches the 0.25-mm$^2$ footprint of MRM-based ones. The bit energy of 0.78 pJ/bit for the driver and modulator is comparable to other state-of-the-art reports [7,10,17]. Given that our design offers scalability in operating speed and is free from temperature control and trimming requirements, our approach is more promising compared with MRM-based transmitters. In the future, the footprint is expected to be further reduced by vertically stacking the driver and modulator through flip-chip integration. Increasing the total data capacity requires large-scale multi-channel integration, which is facilitated by the small footprint PCW MZM. Challenges such as crosstalk that may arise in this process are expected to be mitigated through advancements in co-simulation. Furthermore, integrating the open-collector current-mode driver into a CMOS chip could further improve power efficiency and reduce the footprint.

## 5. Conclusion

We demonstrated a 64-Gbaud optical transmitter consisting of a PCW MZM and open collector driver with a power consumption of

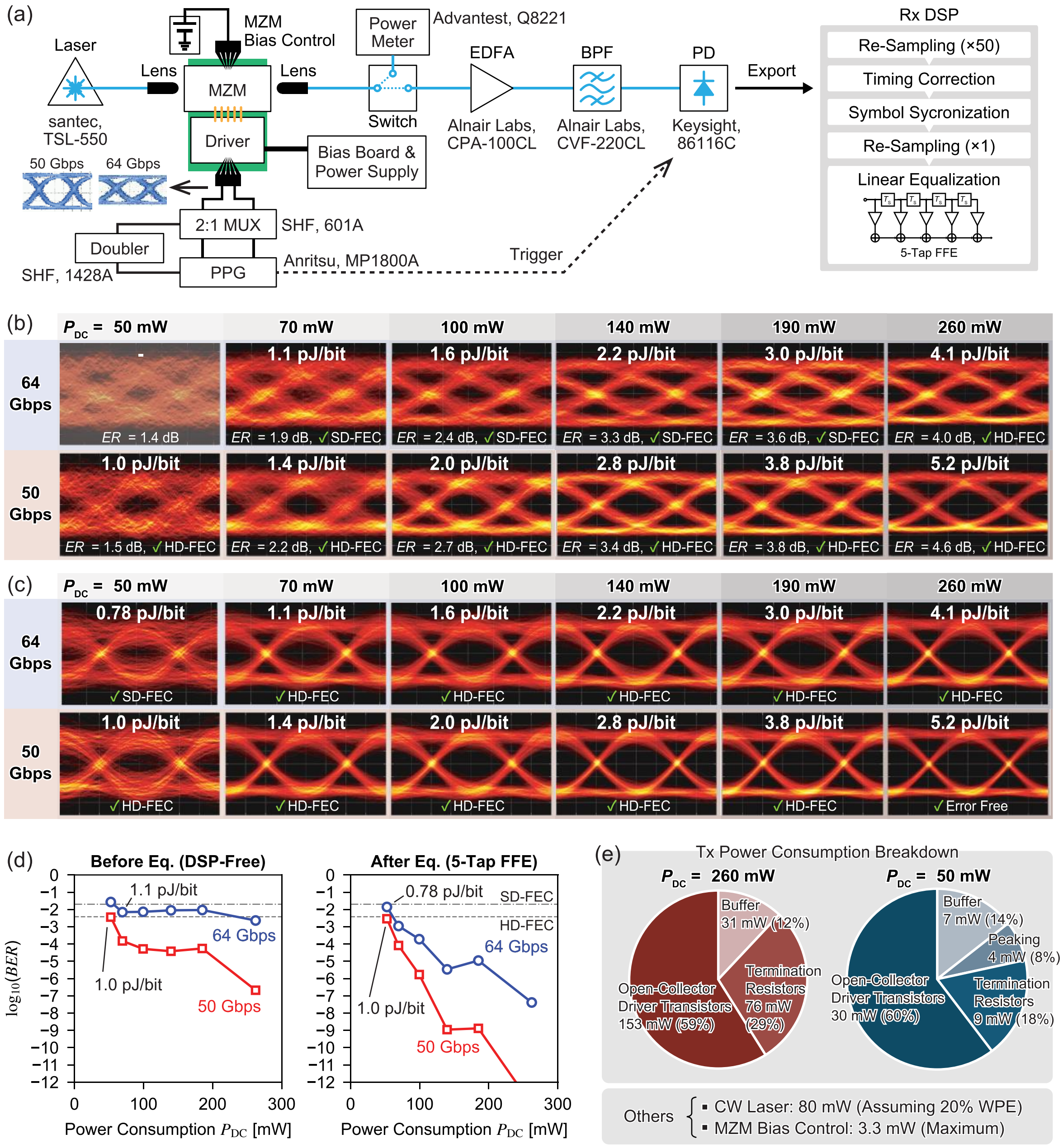

Fig. 5. Signal transmission experiments. (a) Block diagram of the experimental setup. (b) Unprocessed eye diagrams at 50 and 64 Gbps for various power conditions from 50 to 260 mW. (c) Equalized eye diagrams for demonstrating low power transmission. (d) Estimated bit error rate versus the power consumption. (e) Transmitter power consumption breakdown.

50 mW, a bit energy of 0.78 pJ/bit at 64 Gbps, and a footprint of 0.66 mm$^2$. This performance is attributed to the unterminated current-mode driving of compact and high-impedance PCW MZM, where an electronics-photonics co-simulation enabled the prediction of performance and the design of a complicated current-driving circuit. The packaging and design-level integration of photonics and electronics will boost the performance over the traditional separate design approach, which addresses the power reduction issue in optical interconnects.

## Appendix A: Modeling and Design

We built optical device models using Verilog-A language that can be imported into standard electronic design environments. The difficulty in introducing optical devices into circuit simulation is the extremely high frequency of light. Equivalent baseband simulation was used to reduce the time step [28]. The parameters of the optical waveguide and the p-n doped phase shifter were extracted using Lumerical MODE and CHARGE, respectively. The electronics models were provided by the foundry (IHP) and imported into ADS, and the electronics-photonics co-simulation was performed together with the in-house photonic models. Mask layouts for the driver and modulator were designed using KLayout and Cadence Virtuoso, respectively. The wiring parasitics associated with the layout were extracted using Keysight's Momentum electromagnetic field solver and included in the simulations.

Table 1. Comparison between the state-of-the-art integrated optical transmitters.

| Ref. | Photonics Device | Electronics Device | Bandwidth [GHz] | Symbol Rate [Gbaud] | DSP | DC Power [mW] | Bit Energy [pJ/bit] | Footprint† [mm$^2$/Ch.] (Integ. Scheme) |
|---|---|---|---|---|---|---|---|---|
| **This Work** | **Si PCW MZM** | **130-nm BiCMOS** | **55** | **64** (NRZ) | 5-tap Rx FFE | **50** (Driver + MZM) | **0.78** | **0.66** (Wire) |
| [17] | Si Rib MZM | 28-nm HPC+ CMOS | 43 | 112 (NRZ) | 6-tap Rx FFE | 78 (Driver + MZM) | 0.7 | 2.0 (Flip Chip) |
| [15] | Si Rib MZM | 65-nm CMOS | N.R. | 50 (NRZ) | 2-tap Tx FFE | 267.6 (Driver + MZM) | 5.35 | 6.5 (Wire) |
| [14] | Si Rib MZM | 250-nm BiCMOS | 24 | 60 (NRZ) 40 (PAM4) | None | 1770 (Driver + MZM) | 22 | 5.06 (Monolithic) |
| [19] | Si Rib MZM | 250-nm BiCMOS | 35 | 44 (NRZ) | None | 500 (Driver + MZM) | 11.4 | 7.8 (Monolithic) |
| [22] | Si Rib MZM | 55-nm BiCMOS | N.R. | 26.5 (PAM4) | 5-tap Rx FFE | 290 (Driver + MZM) | 5.5 | 11†† (Flip Chip) |
| [37] | SiGe EAM | 55-nm BiCMOS | N.R. | 53 (PAM4) | None | 159 (Driver + EAM) | 1.5 | 0.8 (Wire) |
| [7] | Si Rib MRM | 28-nm CMOS | N.R. | 50 (NRZ) | None | 65 (Driver + MRM + Tuning) | 1.3 | 0.25 (Flip Chip) |
| [10] | Si Rib MRM | 14-nm FinFET CMOS | N.R. | 40 (NRZ) | None | 35 (Driver + MRM + Tuning) | 0.88 | N.R. (Flip Chip) |

†Active footprint including the modulator, driver, and interconnects between the photonics and electronics, which is estimated from the micrograph.

††Including clock data recovery circuits.

## Appendix B: Electronics-Photonics Integration

The modulator and driver were die-bonded on an FR4 printed circuit board (PCB). The PCB was fixed on a base machined from aluminum alloy (A5052). The DC wiring was laid on the PCB to supply power and bias to the driver. The bonding pads on the PCB have a gold flash surface treatment. The chips were attached using conductive epoxy adhesive (CR-2800) by heating at 90°C for 60 min in an oven (KM-300B). The driver and modulator were connected with 20-µm diameter aluminum wires using West Bond's 7476D. The bonding pitch and gap were 100 µm and 400 µm, respectively. To shorten the bonding wire for the RF signal path, the PCB under the modulator was drilled to create a cavity.

## Appendix C: EO Response Measurement

The EO frequency response of the entire transmitter, including the driver, modulator, and bonding wires, was measured using a signal generator (Rohde & Schwarz SMA100B) and an optical sampling oscilloscope (Keysight 86100C + 86116C). Figure 4(a) shows the block diagram of the experimental setup and probing details. The modulator bias was set at the quadrature point. A sinusoidal signal ranging from 1 GHz to 40 GHz was applied to one arm in 1 GHz steps. The amplitude of the sinusoidal signal was set to $V_{pp}$ = 100 mV. The optical filter bandwidth was set to 10 nm to prevent roll-off within the measurement range. To mitigate the impact of measurement noise, waveform averaging was performed 100 times. After the measurement, the electrical frequency response at the cable end was measured to de-embed cable losses. Probe losses were de-embedded using the S-parameters provided by the manufacturer (Technoprobe).

## Appendix D: Signal Transmission Experiments

The performance of the transmitter was evaluated through signal transmission experiments. Figure 5(a) shows the block diagram of the experimental setup. The continuous wave laser light of $P_{LD}$ = 13 dBm was input into the MZM from the tunable laser (Santec, TSL-550). The light was input and output via edge couplers using fiber-lens modules. The output power from the modulator was monitored using a power meter (Advantest, Q8221). The modulator bias was slightly detuned from the quadrature point to increase the *ER*. Two pseudo-random NRZ signals were generated by a pulse pattern generator (Anritsu, MP1800A) and multiplexed to 50 or 64 Gbps using a multiplexer (SHF, 601A). The source signal amplitude was set to $V_{pp}$ = 400 mV. The RF signals were applied directly to the driver via the probe. The modulated light was amplified by an erbium-doped fiber amplifier (Alnair Labs, CPA-100-CL), and the amplified spontaneous emission was removed by a bandpass filter (Alnair Labs, CVF-200-CL) with a bandwidth of $\Delta\lambda$ = 1 nm. The modulated waveform was observed by the optical sampling oscilloscope. After the experiments, DSP including symbol synchronization and 5-tap linear feed-forward equalization was applied at the receiver using MATLAB/Simulink, and no DSP was applied at the transmitter. The BER was estimated by fitting the vertical histogram at the decision point of the eye diagram with two

Gaussian functions.

**Funding.** Japan Society for the Promotion of Science #JP23KJ0988 and #25H00852.

**Acknowledgments.** This work was supported through the activities of VDEC, The University of Tokyo, in collaboration with Cadence Design Systems, Inc. and Keysight Technologies, Inc.

**Disclosures.** The authors declare no conflicts of interest.

**Data availability.** The data underlying the results presented in this paper are not publicly available at this time but may be obtained from the authors upon reasonable request.

**Supplemental document.** See Supplemental Document for supporting content.